\documentclass[5p, times]{elsarticle}
\usepackage{amsmath}
\usepackage{array}
\biboptions{sort&compress, comma}

\date{July 2022}
\journal{Chemical Physics Letters}

\begin{document}
\begin{frontmatter}
\title{Using Machine Learning Hamiltonians To Compute Molecular Motor Barrier Heights}
\author[inst1]{Aaron Philip}
\ead{philipaa@msu.edu}
\author[inst1,inst3]{Guoqing Zhou}
\ead{gzhou@lanl.gov}
\author[inst1,inst3]{Benjamin Nebgen\corref{cor1}}
\ead{bnebgen@lanl.gov}

\cortext[cor1]{Corresponding author}

\address[inst1]{Theoretical Division, Los Alamos National Laboratory, Los Alamos NM, 87544, USA}
\address[inst3]{Center for Integrated Nanotechnologies, Los Alamos National Laboratory, Los Alamos NM 87544, USA}

\begin{abstract}
  Machine Learning Inter-atomic Potentials (MLIPs) have become a common tool in use by computational chemists due to their combination of accuracy and speed. Yet, it  is still not clear how well these tools behave at or near transitions states found in complex molecules. Here we investigate the applicability of MLIPs in evaluating the transition barrier of two, complex, molecular motor systems: a 1st generation Feringa motor \cite{Feringa1999} and the 9c alkene 2nd generation Feringa motor \cite{Vicario2006}. We compared paths generated with the Hierarchically Interacting Particle Neural Network (HIP-NN), the PM3 semi-empirical quantum method (SEQM), PM3 interfaced with HIP-NN (SEQM+HIP-NN), and Density Functional Theory calculations. We found that using SEQM+HIP-NN to generate cheap, realistic pathway guesses then refining the intermediates with DFT allowed us to cheaply find realistic reaction paths and energy barriers matching experiment, providing evidence that deep learning can be used for high precision computational tasks such as transition path sampling while also suggesting potential application to high throughput screening.
\end{abstract}
\end{frontmatter}

\section*{Introduction}
Since the awarding of the 2016 Nobel Prize in Chemistry, molecular motors and other molecular-mechanical devices have been a huge interest due to their application to nanomachinery and drug synthesis.
In nature, common examples of molecular motors are found in the ATP synthase enzyme in cells that synthesize the energy molecule adenosine triphosphate via a rotary motion as well as the kinesin proteins that utilize molecular motors to "walk" along microtubules to transport cellular products. As such, studies of molecular motors, switches, gears, propellers and many more mechanical analogues on the nanoscale allow computational chemistry to inform related disciplines ranging from biology and medicine to nanorobotics. By computationally modeling the behavior of such mechanical molecules, tasks such as high throughput screening can be optimized for experimental application. 

However, a fundamental problem in computational chemistry today is finding methods that model chemical properties while striking a reasonable balance between computational cost and accuracy. \textit{Ab initio} methods such as Coupled Cluster \cite{Cizek1966, Bartlett2007} or Density Functional Theory (DFT) \cite{Kohn1965} are popular choices for generating accurate solutions, but scale poorly in cost with increasing electron number. Alternatively, tight binding \cite{Porezag1995, Elstner1998} or semi-empirical quantum methods are more efficient that come at the cost of reduced accuracy or generality. These methods utilize empirical parameters fit to either experimental data or quantum calculations to reduce the number electrons in a given  simulation. Although successful at accelerating molecular dynamics simulations, the methods often transfer poorly to accurately describing other systems. With recent advances in machine learning (ML) for scientific computing, chemists are now able to find more favorable compromises between accuracy and efficiency. ML models are capable of automatically constructing potentials with training times that scale linearly to datapoints. Please see recent review articles for more details \cite{Kulichenko2021, Zubatiuk2021, Dral2020}.

The promise of ML in chemical modeling allows previously costly simulations of system dynamics to become far cheaper, opening the door for simulations of more complex systems. Transition path sampling focuses on accurately modeling the mechanisms by which molecular systems undergo rare transitions between stable states. Sampling can be done by identifying saddle points on the energy landscape and examining paths that cross these states. However with complex landscapes with many degrees of freedom, there may be far too many saddle points to examine, making the task of identifying important transition states too difficult and costly. Common sampling practice involves Monte Carlo shooting and shifting schemes \cite{Dellago1998, Bolhuis1998, Metropolis1953, Crooks2001} that require only specifying endpoint states and constructing an ensemble of random walks weighted with appropriate probabilities so that probable transition states are explored more frequently. Due to the expensive nature of path sampling, it is of utmost importance to calculate atomic and molecular energies with high accuracy and low cost during the sampling process.

In this paper we examine the applicability of a variety of ML, semi-empirical, and full quantum models to correctly compute the transition state for two different molecular motor species. For the pure ML based approach, we use the newly developed HIP-quad tensor architecture \cite{Lubbers2018}. For the pure semi-empirical based approach, we utilize PM3 \cite{Stewart1989}. In addition, we also include the newly released ML modified semi-empirical method, SEQM+HIP-NN \cite{Zhou2022}. Finally, we utilize BP86/def2-XVP
as implemented in ORCA as a reference DFT methodology. In Molecular Motors we discuss the specific molecular systems examined in this paper. In Methods we discuss the various methodologies used to determine transition barriers of the molecular motors. In Results we show how each of these methods compare at finding the transition barrier, compare the optimized geometries they find, and discuss the implications of these results. 

\section*{Molecular Motors}
\label{sec:MMs}
\begin{figure}
    \centering
    \includegraphics[width=3in]{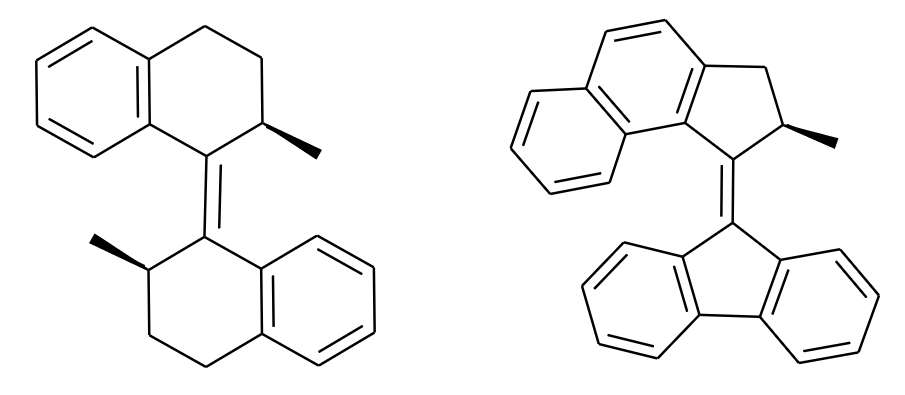}
    \caption{\textit{Left:} One of the 1st Generation Motors. \textit{Right:} One of the 2nd Generation Motors..}
    \label{fig:motors}
\end{figure}
\subsection*{1st Generation Motor}
The 1st generation motors \cite{Feringa1999, Roke2018, Feringa2017, terWiel2005} rotate as a result of alternating photochemical and thermal stimuli. The photochemical steps enable the molecule to under go \textit{cis-trans} isomerizations about the central carbon double bond, while the thermal steps cause helical inversion, preventing the molecule from reverse rotation. In the original molecule, four steps were required to undergo a full 360$^{\circ}$ rotation, with two photochemical steps stimulating \textit{cis-trans} isomerizations and two thermal steps causing helical inversion \cite{Feringa1999}. These trajectories were non-linear in nature, with each side twisting out of the plane whenever they passed by each other. We modeled a related 1st generation molecule shown in Fig. \ref{fig:motors} with no experimental data published. With this molecule, the step most difficult to model was the helical inversion occurring when the two methyl groups slid by one another while in the \textit{cis} configuration. As such, we chose to model this molecule through a transition from a \textit{cis} configuration to a \textit{trans} configuration while undergoing the helical inversion associated with a thermal stimuli--effectively modeling three of the four steps required to complete a full rotation. To transition through these three steps, the experimental barrier height for the original motor was 1.14 eV \cite{Feringa1999}, which we used as a general guide as to where the barrier for the alternate 1st generation motor should be.
\subsection*{2nd Generation Motor}
As shown in Fig. \ref{fig:motors}, the 2nd generation molecular motors \cite{Cnossen2014, Koumura2000, vanDelden2003, Pijper2007} have a rotating side that is symmetric about the carbon=carbon bond it rotates about. As such, energetic barriers were lowered for the transition, and modeling the rotation was reduced to a problem of only modeling a 180 $^\circ$ rotation relative to the other side, because the exact same initial configuration was achieved after the molecule had passed through three of the four stimuli required to achieve a full rotation. The experimental barrier for the entire transition we modeled was recorded as 0.698 eV \cite{Vicario2006}. 
\section*{Methods}
\label{sec:methods}
\subsection*{HIP-NN}
HIP-NN is a neural network architecture capable of predicting molecular properties and energies. Molecules are passed in as a collection of atomic species and coordinates and HIP-NN creates a representation of each atom's chemical environment in order to then make predictions using on-site layers and interaction layers (discussed below). HIP-NN obtains total energy $E$ by considering the total energy of a molecule as an approximate sum of atomic energies $\hat{E_{i}}$ over all atoms within the molecule: 
\begin{equation}
    E = \sum_{i=1}^{N_{atom}} \hat{E_i}
\end{equation}
In this consideration, the atomic energy $\hat{E_i}$ is composed of a sum of hierarchical energy terms (blue boxes in Fig. 1) of order $n = 0$ up to the number of interaction layers in the network $N_{interaction}$ in the form:
\begin{equation}
    \hat{E_i} = \sum_{n=0}^{N_{interaction}} \hat{E_{i}^{n}}
\end{equation}
By tracking the relative $\hat{E_i^{n}}$ magnitudes at each hierarchical level $n$, HIP-NN is able to estimate the reliability of its predictions. Correct energy predictions will typically have rapidly decaying $\hat{E_i^{n}}$ with increasing $n$ in the network. If such decay is not observed (i.e. non-hierarchicality), HIP-NN is able to convey a lack of confidence in its energy prediction.
\begin{figure}[h]
    \centering
    \includegraphics{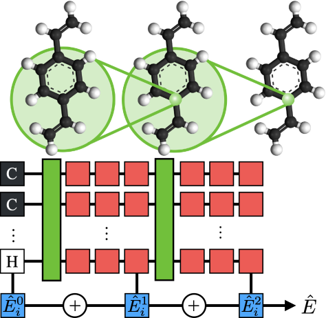}
    \caption{HIP-NN model structure: One-hot encoded input (black and white blocks), interaction layers (green blocks), on-site layers (red blocks), and hierarchical energy terms (blue blocks)}
    \label{fig:my_label}
\end{figure}

As a molecule pass through the network, each composite atom is represented by a set of features characterizing its environment in which a single feature of an atom can be represented as $z^{l}_{i, a}$ where $l$ denotes layer, $i$ denotes atomic index, and $a$ denotes feature index. Molecules enter the network and are subsequently one-hot encoded by atomic species which is represented by the 2-D array $z^{0}$. As the molecule representation passes through the network, on-site layers build features for each individual atom while interaction layers account for atom-atom interactions and pass information between atoms that fall within a certain locality.

On-site layers (red boxes in Fig. 1) act as operators on individual atoms' features by taking in an atom's features $z^{l}_{i}$ and outputting transformed features $z^{l+1}_{i}$ according to the rule:
\begin{equation}
    z^{l+1}_{i, a} = f(\sum_{b}W_{a,b}^{l}z_{i,b}^{l} + B_{a}^{l})
\end{equation}
Here, $f(x)$ is the softplus activation function $log(1+e^x)$, $b$ is the atomic feature index in layer $l$, $a$ is the atomic feature index in layer $l+1$ , $W^{l}$ is a tensor of learnable parameters, and $B^{l}$ is a bias vector of learnable parameters.

Interaction layers (green boxes in Fig. 1) also behave as operators with the additional function of passing information between nearby atoms. The coordinates of each atom in the molecule are used to encode the pairwise distances between atoms $r_{ij} = |r_{i}-r_{j}|$ within a certain locality such that $r_{ij} < R_{cut}$ where $R_{cut}$ is a radial threshold with a smooth cutoff. This allows the network to learn from the relative positions of atoms rather than converging to different predictions when the molecule is translated, rotated, or reflected in space which can occur when absolute atomic coordinates are used \cite{Behler2007}. Interaction layers operate on previous layers with the following form:
\begin{equation}
    z^{l+1}_{i, a} = f(\sum_{j,b}v^{l}_{a,b}(r_{ij})z^{l}_{j,b} + \sum_{b}W^{l}_{a,b}z^{l}_{i,b} + B^{l}_{a})
\end{equation}
Here, $j$ refers to other atoms around atom $i$ (green circles in Fig. 1) and $v^{l}_{a,b}(r_{ij})$ regulates the influence of a nearby atom $j$ based on the pairwise distance $r_{ij}$ from atom $i$. This is done according to: 
\begin{equation}
   v^{l}_{a,b}(r_{ij}) = \sum_{v}V^{l}_{v,a,b}s^{l}_{v}(r_{ij}) 
\end{equation}
$V^{l}_{v,a,b}$ are learnable parameters and $s^{l}_{v}(r_{ij)}$ is a sensitivity function given by:
\begin{equation}
   s^{l}_{v}(r) = \exp[-\frac{(r^{-1}-\mu_{v,l}^{-1})^2}{2\sigma_{v,l}^{-2}}]\varphi_{cut}(r) 
\end{equation}
$\mu_{v,l}$ and $\sigma_{v,l}$ are additional learnable parameters while the smooth cutoff function $\varphi_{cut}(r)$ is defined as:
\begin{equation}
    \varphi_{cut}(r) = \begin{cases}
    \cos^2(\frac{\pi}{2}\frac{r}{R_{cut}}) 
    & r_{ij} \leq R_{cut },\\
    0 & r_{ij} > R_{cut}.
    \end{cases}
\end{equation}
After the molecule is processed throughout the entire HIP-NN structure, the hierarchical energy terms for each atom $\hat E_i^n$ are summed to obtain individual energies for each atom $\hat E_i$. In turn, total molecular energy $E$ is obtained by summing atomic energies $\hat E_i$ over each atom in the molecule. 

\subsection*{PM3}
For a semi-empirical model, we utilize the PM3 semi-empirical Hamiltonian\cite{Stewart1989} as implemented in the PYSEQM software package \cite{Zhou2020}. The PM3 method assumes that the total energy can be expressed in terms of electronic energy and nuclear energy. Electronic energy is in turn represented by one-electron core Hamiltonians that are parametrized, with each atomic element having a corresponding 18 parameters except for hydrogen which has 7 parameters. Nuclear energy is likewise parameterized. PM3 uses these core hamiltonians and the nuclear energy to then generate the complete Hamiltonian using SCF procedure. The PM3 method relies on obtaining optimal parameter sets by fitting these parameters to minimize reference functions describing the discrepancy between PM3 calculation of molecular properties and reference experimental results or even high-level \textit{ab initio} calculations. 

\subsection*{DFT}
For the DFT methdology we utilize the BP86 level of theory as implemented in the ORCA software package \cite{Neese2020} in conjunction with the def2-XVP basis sets. The method uses Density Functional Theory in which one-electron Kohn-Sham equations are utilized alongside interelectronic interactions to model the electric behavior of the system by using electron density to construct effective potentials. BP86 then utilizes revised functionals to refine exchange and correlation energy approximations. Exchange energy utilizes gradient-based corrections to achieve an approximation with correct asymptotic potential in large systems using a one-parameter functional \cite{Becke1988}. The correlation energy term is likewise obtained using a functional providing strong approximations to experimental results, while also extending robustly to slowly varying electron gas densities \cite{Perdew1986}.

\subsection*{SEQM + HIP-NN}
The PYSEQM module \cite{Zhou2020} interfaced with HIP-NN has proven to regularly surpass the energy prediction accuracy of HIP-NN on organic molecules. PYSEQM utilizes the HIP-NN architecture to generate dynamic parameters for a Hamiltonian that is then iteratively refined according to SCF procedure using the PM3 method to obtain a density matrix and final Hamiltonian which are then used to calculate electronic, nuclear, and correction energy terms \cite{Zhou2022}. The entire model was trained via backpropogation through the structure against a subselection of the ANI-1x dataset \cite{Smith2018} of Density Functional Theory calculations on small organic molecules. 

\subsection*{Path Generation}
Both molecules being studied achieved their motor function by rotating about the central carbon-carbon bond which behaved as an "axle". To define the state of the motors, we used this axle and two directly connected carbon atoms to compose a dihedral angle. The difference between the dihedral at the initial and final geometries was used to guide the molecule through its transition. If $C_i$ and $C_f$ are the set of coordinates of initial and final structures respectively, and $D(C)$ is the operator that returns the value of the dihedral angle for a set of coordinates, then 
\begin{equation*}
    \theta_{i} = D(C_i)
\end{equation*}
\begin{equation*}
    \theta_{f} = D(C_f)
\end{equation*}
\begin{equation}
    \Delta\theta = \theta_{f} - \theta_{i}
\end{equation}
where $\theta_{i}$ and $\theta_{f}$ are the dihedral angles of the initial and final states respectively.
The dihedral angles of intermediate states are subsequently generated by
\begin{equation}
    \theta_{i+1} = \theta_{i} + \alpha\Delta\theta
\end{equation}
such that if $I$ is the desired number of intermediates, $\alpha$ is
\begin{equation}
    \alpha = (I+1)^{-1} 
\end{equation}

Then if $C_i$ is defined as the set of coordinates composing an intermediate, then each subsequent set of coordinates $C_{i+1}$ is generated from the previous set $C_{i}$ by imposing the constraint that
\begin{equation}
    D(C_{i+1}) = \theta_{i+1}
\end{equation}

We used the trust region constrained optimization algorithms \cite{Conn2000} in the SciPy package alongside HIPNN and PYSEQM to find $C_{i+1}$ from $C_i$ by shifting $C_i$ to find a local energy minima that satisfies the dihedral constraint. Because PYSEQM and HIPNN are built with PyTorch, automatic differentiation enabled calculations of gradients of energy with respect to position. Conservative forces are defined by
\begin{equation*}
    F =  - \nabla{E}
\end{equation*}
so whenever one of the model structures generated an energy prediction, the gradients of the energy with respect to positions allowed us to obtain forces behaving on each atom. The optimization algorithms used the jacobian to find minima on the energy landscape, physically translating to pushing atoms in the direction of the net forces acting on them.

In order to generate a path from scratch, we used the \linebreak SEQM+HIP-NN model with two adjacent dihedral angle constraints to force the molecule through the transition while optimizing, then optimized each intermediate again with only the guiding dihedral angle held as a constraint.  

After an initial trajectory was generated, we allowed the pure HIP-NN model and pure PM3 method optimize and evaluate the energy of the intermediates on the existing path. Additionally, we ran DFT constrained geometry optimization using the ORCA quantum chemistry package \cite{Neese2020}.
\section*{Results and Discussion}
\label{sec:results}
\subsection*{Energy Surfaces}
\begin{figure*}
    \centering
\includegraphics[scale=0.5]{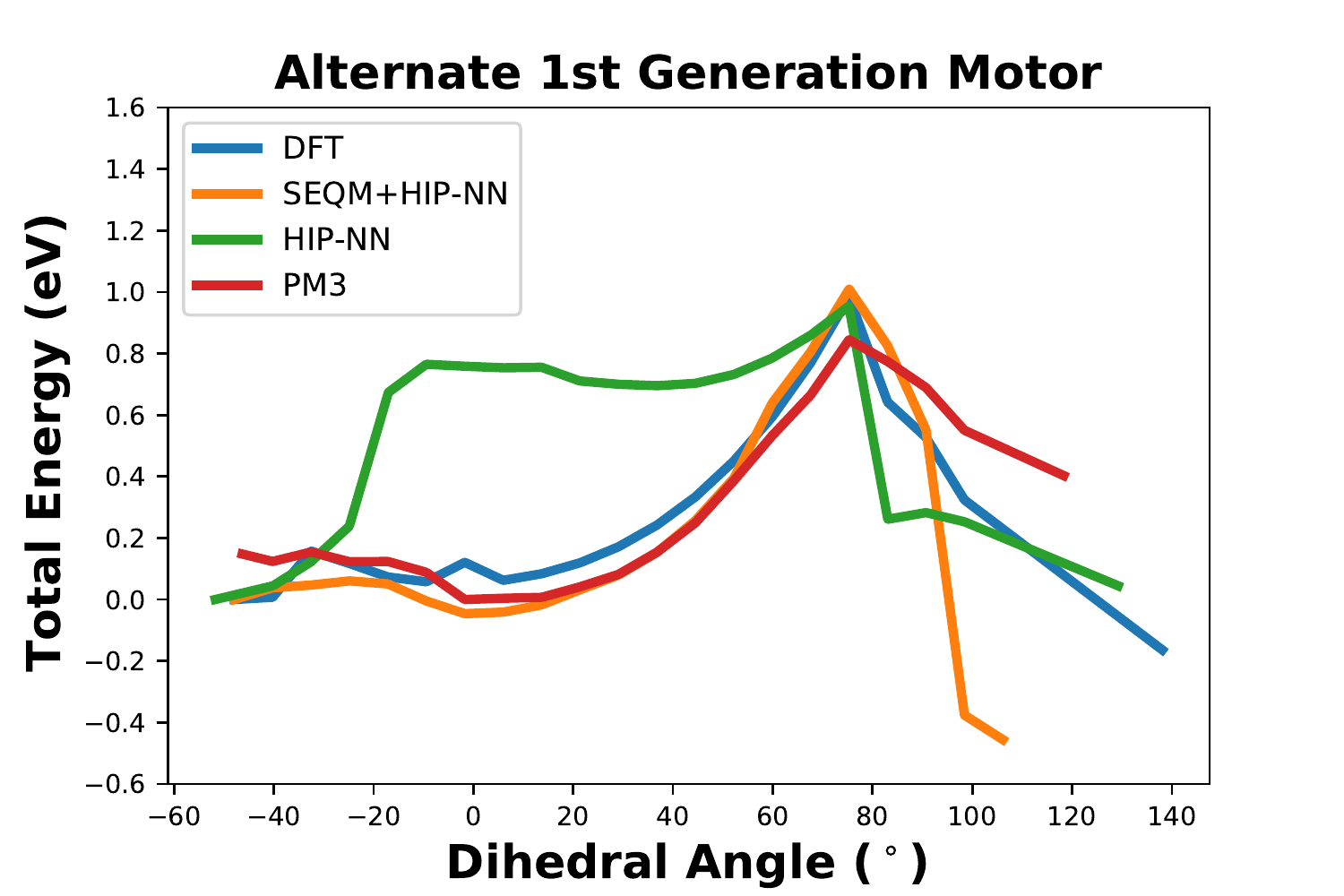}
\includegraphics[scale=0.5]{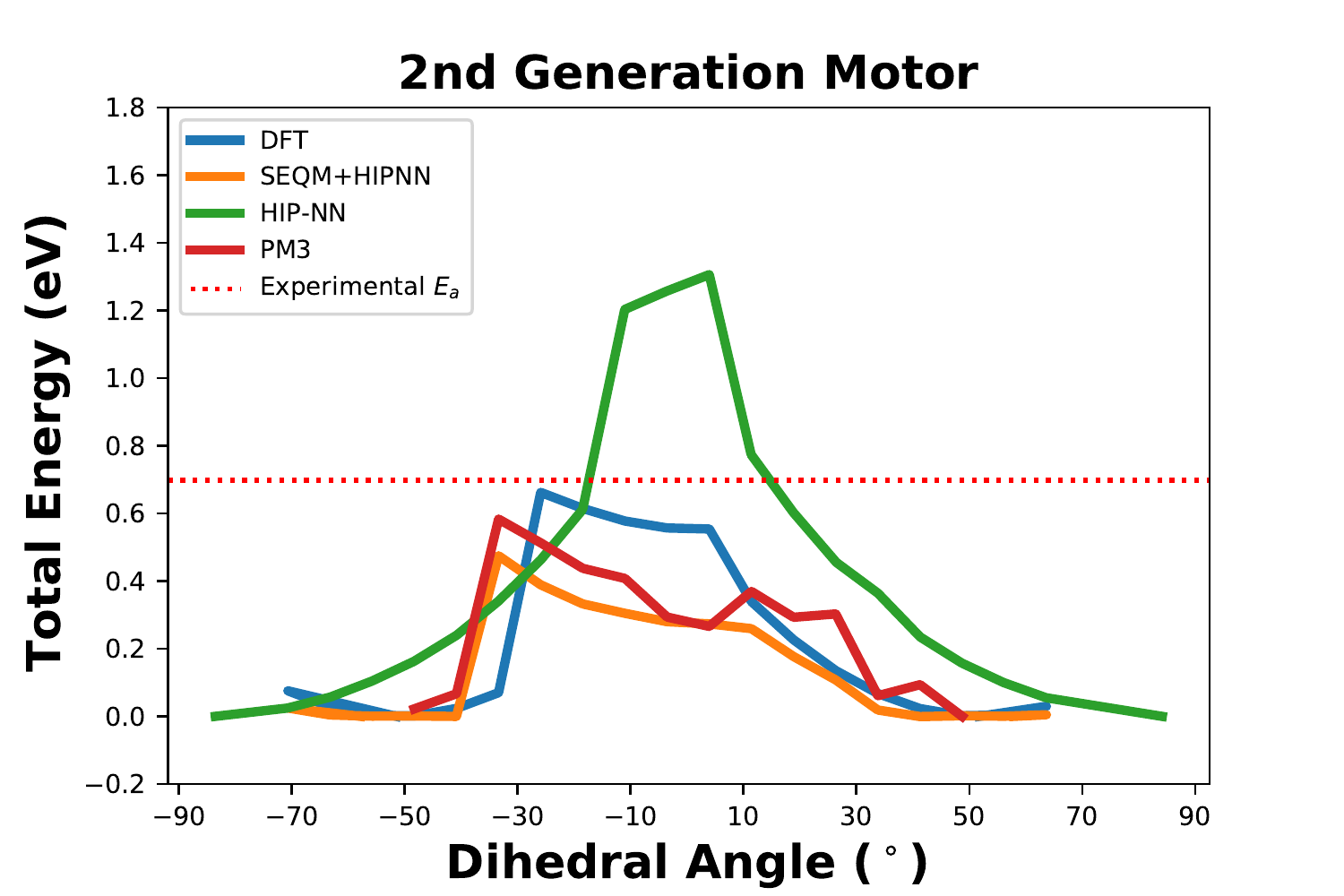}
    \caption{Energy profiles for geometries on initial trajectory of the 1st generation molecular motor and the 2nd generation motor optimized with dihedral angle at each intermediate.}
    \label{fig:surfaces}
\end{figure*}
\begin{figure*}[]
    \centering.
    \begin{center}
    \begin{tabular}{ccc}
    \small\bf  Initial State &
    \small\bf Transition State &
    \small\bf Final State \\
    \hline
    \includegraphics[width=2.in]{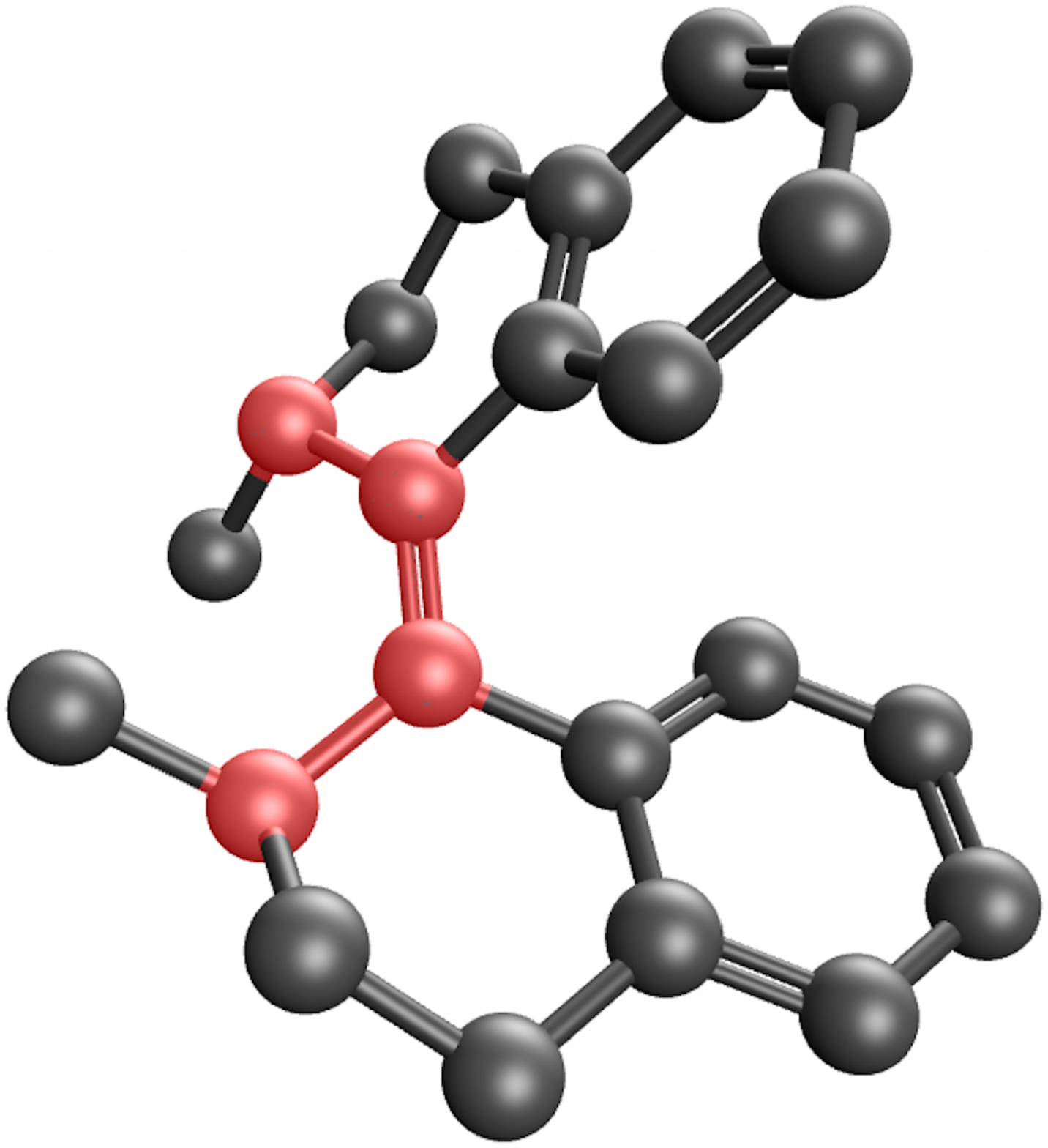} & \includegraphics[width=2.in]{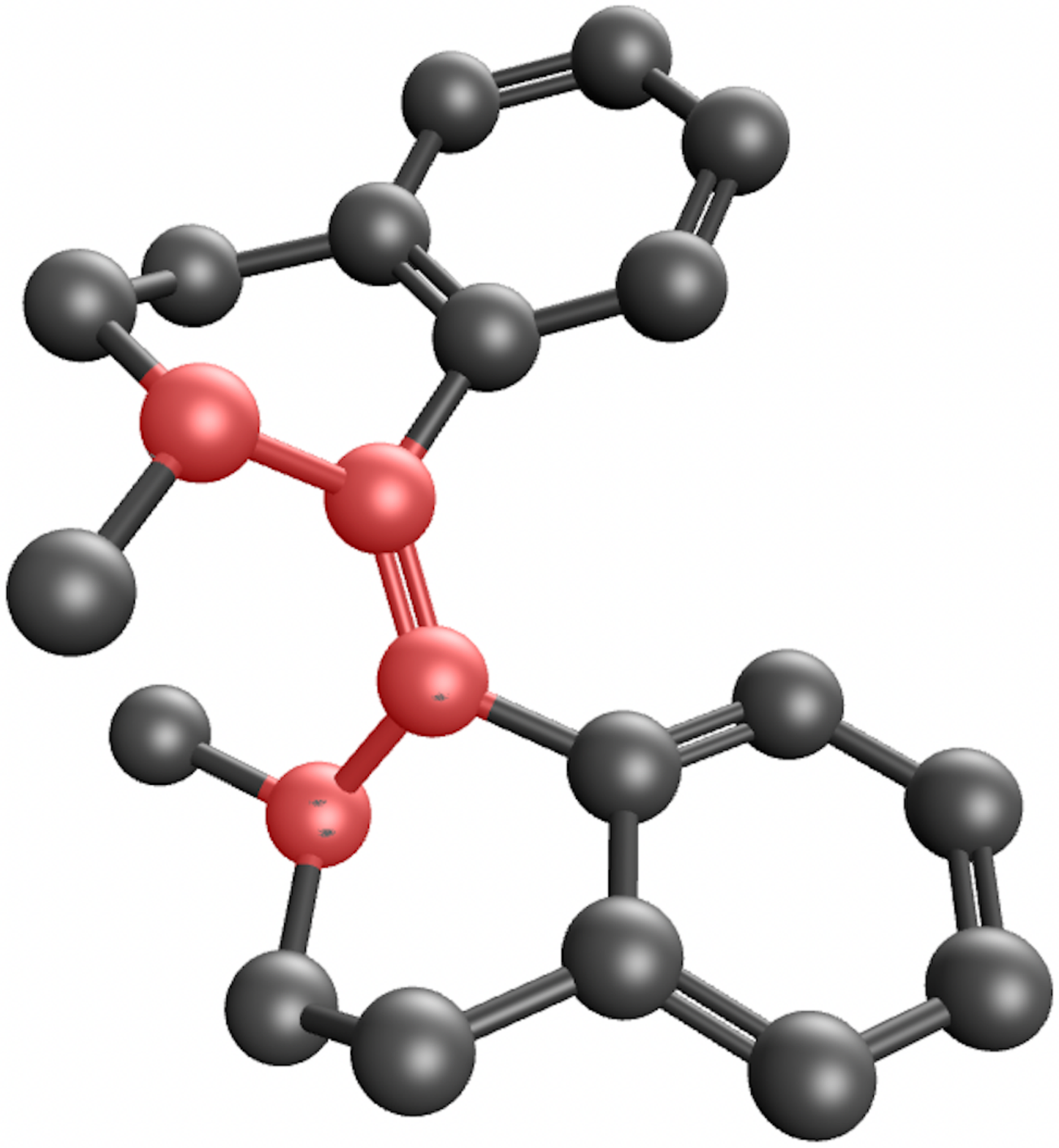} &  \includegraphics[width=2.in]{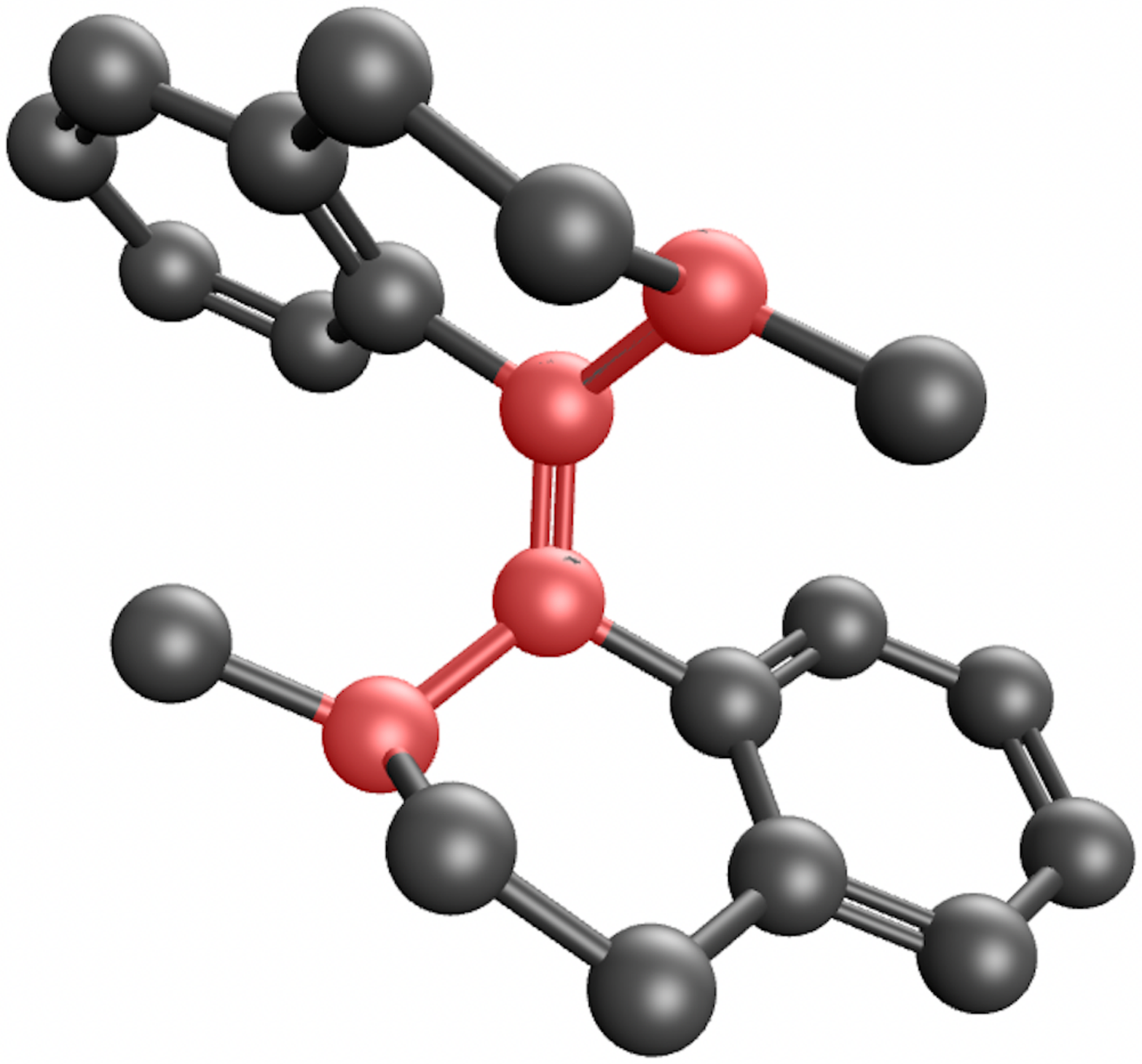} \\ 
    \hline
    \includegraphics[width=2.in]{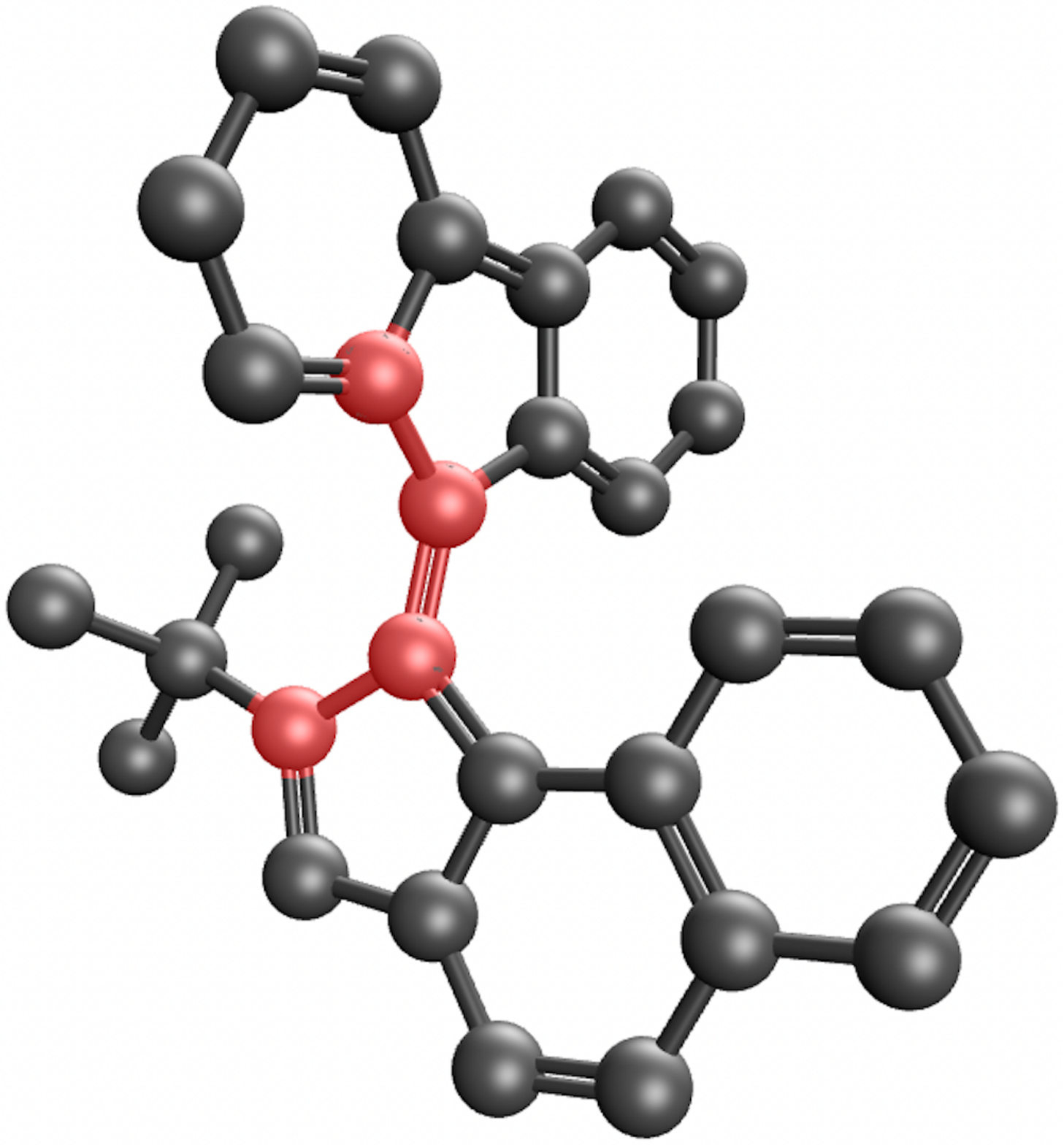} & \includegraphics[width=2.in]{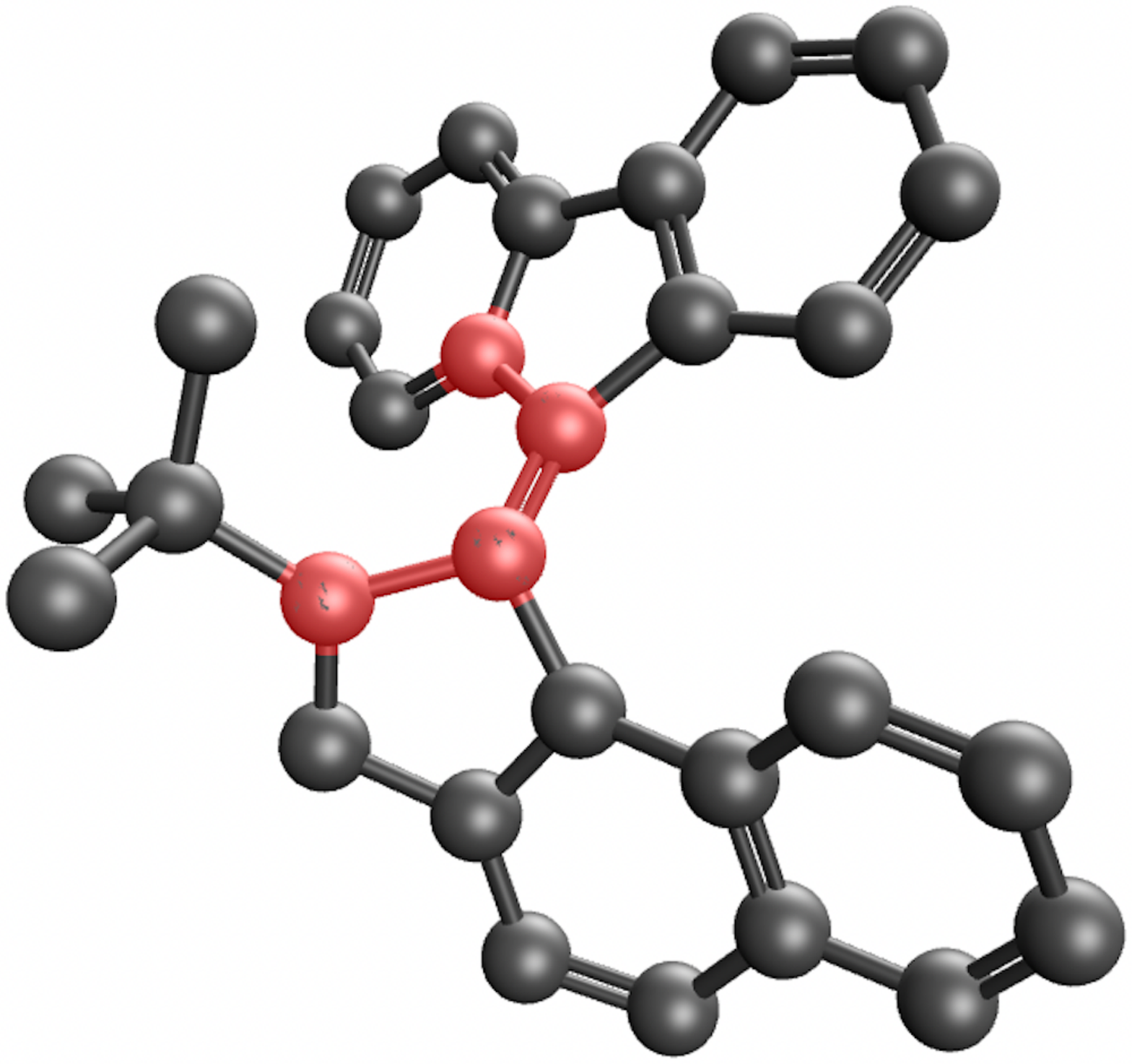} & \includegraphics[width=2.in]{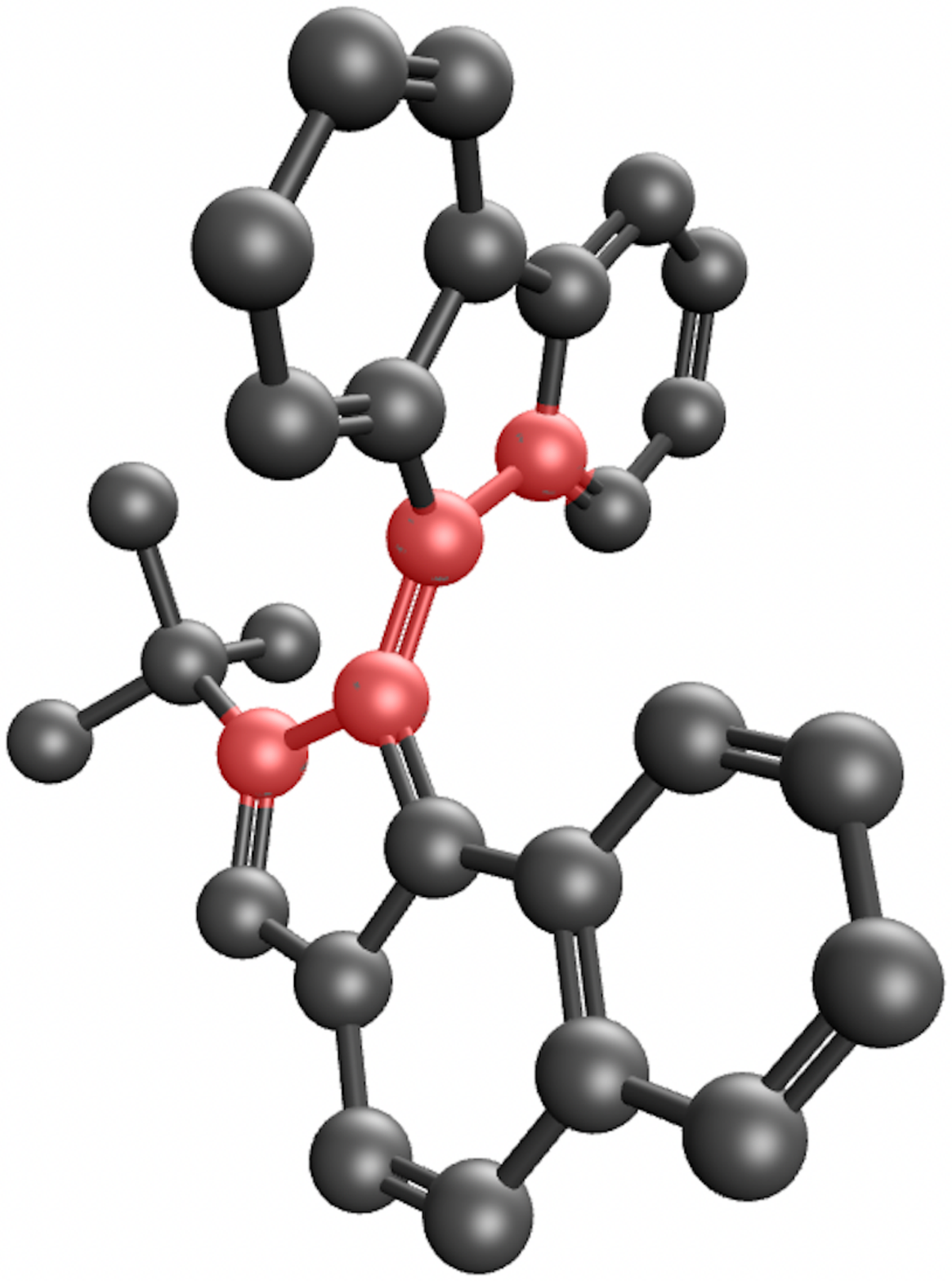} \\   
    \end{tabular}
    \end{center}
    \caption{\textit{Top Row:} 1st Generation Molecular Motor. \textit{Bottom Row:} 2nd Generation Molecular Motor. Red atoms in each row indicate the atoms composing the dihedral angle which was used as the collective variable to guide the transition. Only the carbon backbone is shown; hydrogen atoms have been omitted for visualization purposes.}
    \label{fig:visualize}
\end{figure*}
Fig \ref{fig:surfaces} gives a comparison of HIPNN, PM3, SEQM+HIPNN, and DFT and their relative energy evaluations and optimizations. For the 1st generation molecular motor, each method identified approximately the same energetic saddle point at which the transition state occurred. Although HIP-NN was able to evaluate the endpoint state energies relatively accurately, it suffered when attempting to predict energies of intermediates. This suggests that for this molecule, HIP-NN is still unstable at evaluating non-equilibrium configurations. SEQM+HIP-NN generated a very continuous reaction path but failed to correctly evaluate the energy of the final state whereas the pure PM3 method more or less traced out the SEQM+HIP-NN energy profile until it attempted to optimize geometries past the transition state. The experimental activation energy for the original 1st generation motor in \cite{Feringa1999}) was 1.14 eV and all four methods found paths with similar activation energies that are slightly lower than 1.14 eV as can be expected due to the omission of a third ring on each side of the molecule: HIP-NN peaked at 0.95 eV, SEQM+HIP-NN at 1.01 eV, PM3 at 0.86, and DFT at 0.98 eV.

The same procedure yielded very different results when applied to the 2nd generation molecular motor. DFT and the SEQM+HIP-NN model identified different dihedral angles than the HIP-NN model at which the transition state occurred. Each method also identified different barrier heights for the reaction, with the SEQM+HIPNN model identifying the lowest energy path for the transition. The experimental energetic barrier reported in \cite{Vicario2006} was 0.698 eV and the path optimized with DFT was closest with a barrier height of 0.662 eV. All four methods identified endpoints that were essentially the same energy as each other when allowed to converge the energy with unconstrained optimization. These initial and final endpoints had the same energy as a result of all three methods identifying an optimized final geometry that is a spatial reflection of the optimized initial geometry. 

Generating an initial optimized path with the SEQM+HIP-NN model required between 2-5 hours of computation time on a single CPU node to generate 19 intermediates in addition to optimizing endpoint structures. HIP-NN required ~30 minutes to optimize all structures on the generated path. For reference, pure PM3 took between 15-30 minutes per geometry and optimization jobs on ORCA required between 1-2 hours on a single node to optimize each geometry. An additional consideration is that PYSEQM has GPU capabilities that were not utilized for most of the study and considerably accelerated calculation times and SCF convergences. The energy surfaces of the transition paths generated with each method are displayed in Fig. \ref{fig:surfaces}, while visuals of the initial, final, and transition state geometries are given in Fig. \ref{fig:visualize}.
\subsection*{Path Differences}
\begin{figure}[h]
    \centering
\includegraphics[scale=0.5]{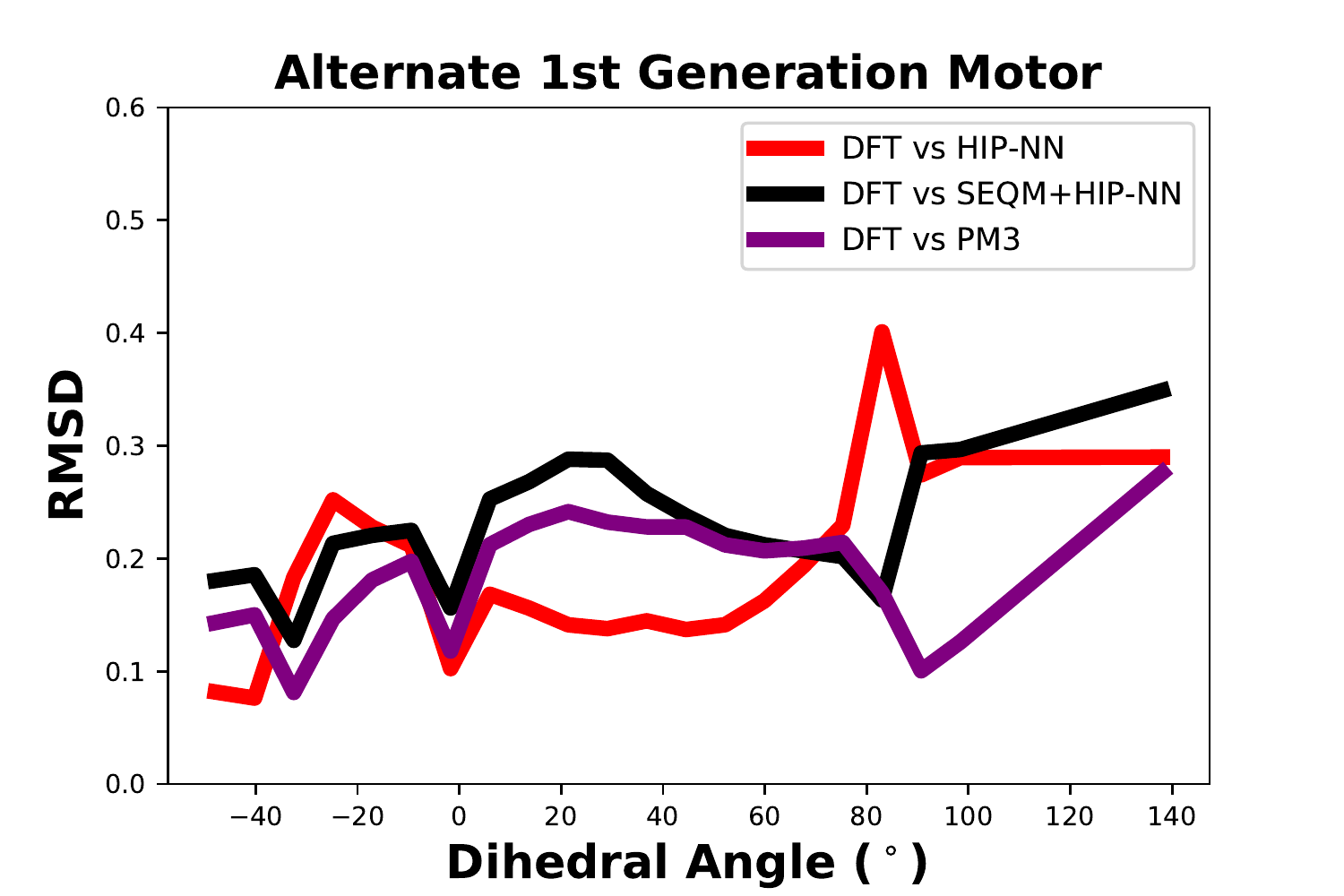}
\includegraphics[scale=0.5]{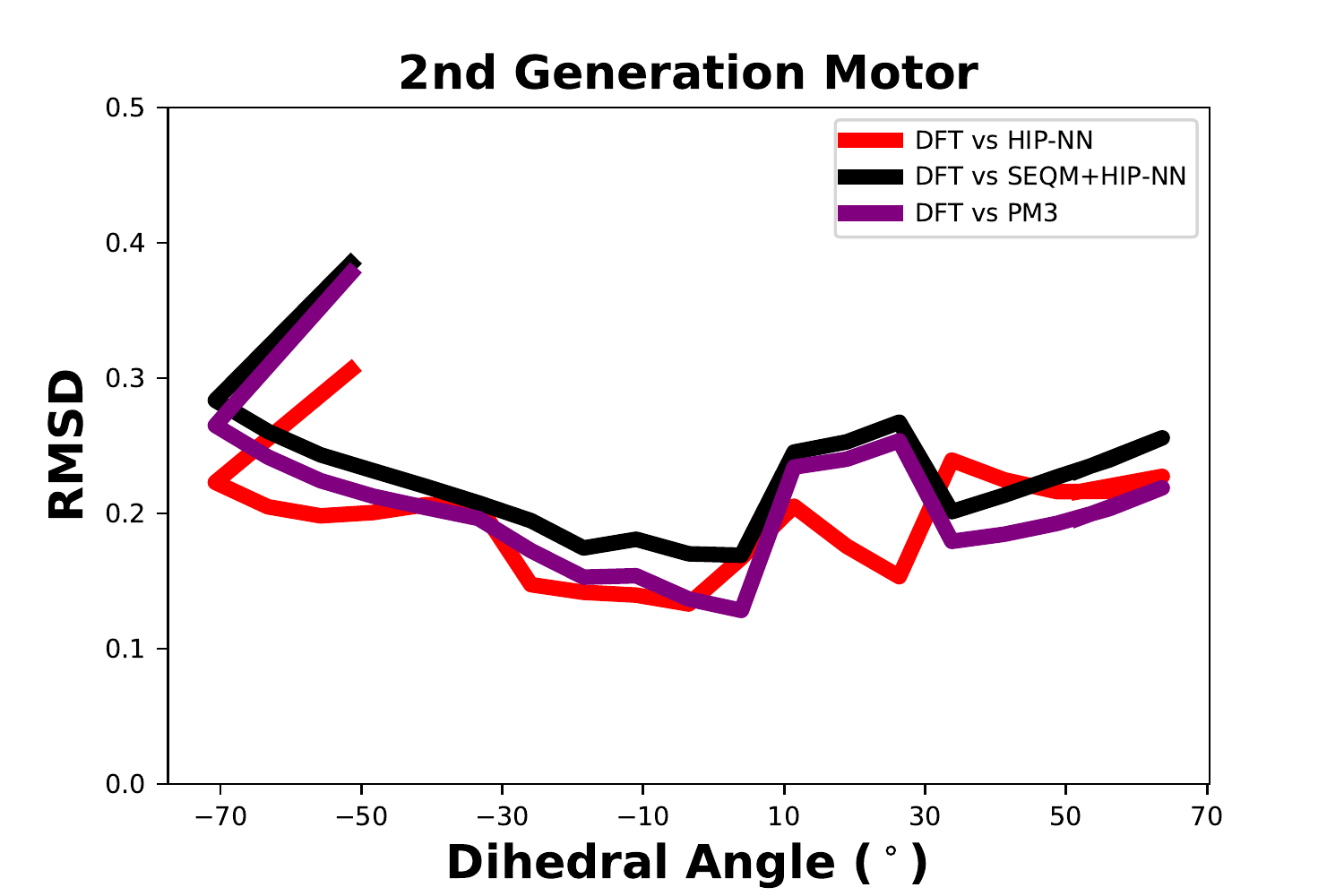}
    \caption{Root mean squared distances between the atoms of geometries optimized with different methods for the 1st generation motor and the 2nd generation motor.}
    \label{fig:RMSE}
\end{figure}
Fig. \ref{fig:RMSE} displays the differences between intermediates generated using each method by measuring the Root Mean Squared Distances between atoms in intermediates generated with each method. As Fig. \ref{fig:surfaces} demonstrates, HIP-NN evaluates several of its generated intermediates as having much higher energies than the other methods. However, Fig. \ref{fig:RMSE} shows that the HIP-NN geometries don't significantly differ from those generated with DFT relative to those generated with other methods. This suggests that HIP-NN is capable of generating realistic non-equilibrium geometries, but is poor at evaluating their energies. After the transition state is crossed, the RMSD spikes however. PM3 and SEQM+HIPNN both optimized to very similar geometries as shown by the RMSD plots in Fig \ref{fig:RMSE} and the energy surfaces in Fig. \ref{fig:surfaces}, although after the transition state in the 1st generation motor the reaction paths began to diverge.
\section*{Conclusions}
In the study of both molecular motors, reasonable transition paths with viable physical geometries were generated using ML methods that greatly accelerated computation time. As discussed, ORCA optimization jobs near a energy minima required ~1-2 hours of computational time on a node using DFT calculations to optimize a geometry generated and optimized with the SEQM+HIP-NN model. HIP-NN required even less time (~30 minutes), but suffered from poor accuracy in evaluating energies. On the contrary, the SEQM+HIP-NN structure achieved strong agreements with DFT in energy evaluations, particularly when modeling the transition path of the 1st generation motor. With the 2nd generation motor, SEQM+HIP-NN chose optimized intermediates that found the transition state at different dihedral angles while still achieving similar energy evaluations to DFT calculations. Particularly noteworthy is the fact that the activation energies discovered in both molecules reflected experimental results, indicating that the paths discovered were strong representations of what actually occurs on the molecular scale.

Albeit not perfectly matched to DFT calculations, SEQM+HIP-NN was able to find strong guesses for intermediates at cheap computational costs that could then be refined with \textit{ab initio} calculations. As such, we have demonstrated that deep learning can be leveraged to accelerate high cost calculations while maintaining agreeable accuracy, even when applied to volatile tasks such as path sampling that can easily find unphysical geometries or destroy the molecule attempting to reach an endpoint state. Computational costs to generate a similar path with purely \textit{ab initio} calculations can be avoided using the SEQM+HIP-NN model to quickly generate reasonable initial guesses and then refine with DFT. Probability-based sampling techniques such as Monte Carlo shooting and shifting schemes may be able to enhance this process by incorporating ML models such as HIP-NN or SEQM+HIP-NN into their explorations of high-dimensional search space; this is one of the many promising future steps forward. 
\section*{CrediT Authorship Contribution Statement}
\textbf{Aaron Philip}: Software, Investigation, Formal Analysis, Data Curation, Visualization,
Writing-Original Draft. \textbf{Guoqing Zhou}: Software, Resources, Supervision. \textbf{Benjamin Nebgen}: Conceptualization, Methodology, Funding Acquisition, Project Administration, Supervision, Writing-Review and Editing. 
\section*{Declaration of Competing Interest}
The authors declare that they have no known competing financial
interests or personal relationships that could have appeared to influence
the work reported in this paper.

\section*{Acknowledgements}
A.P. and G.Z. acknowledge support from the Laboratory Directed Research and Development (LDRD) program for the funding for this work. B.N. acknowledges support from the US DOE, Office of Science, Basic Energy Sciences, Chemical Sciences, Geosciences, and Biosciences Division under Triad National Security, LLC (“Triad”) contract Grant 89233218CNA000001 (FWP: LANLE3F2). This work was performed in part at the Center for Nonlinear Studies and the Center for Integrated Nanotechnology, a US Department of Energy (DOE) and Office of Basic Energy Sciences user facility. We thank Wei Li, Ying Wai Li, and Nicholas Lubbers for useful conversations about the current work and its future directions. This research used resources provided by the LANL Institutional Computing Program, which is supported by the US DOE National Nuclear Security Administration under Contract 89233218CNA000001. We also acknowledge the CCS-7 Darwin cluster at LANL for additional computing resources.
\bibliography{bib}{}

\begin{thebibliography}{10}

\bibitem{Feringa1999}
N.~{Koumura}, R.~W.~J. {Zijlstra}, R.~A. {van Delden}, N.~{Harada}, and B.~L.
  {Feringa}, ``Light-driven monodirectional molecular rotor,'' {\em Nature},
  vol.~401, pp.~152--155, Sept. 1999.

\bibitem{Vicario2006}
J.~Vicario, M.~Walko, A.~Meetsma, and B.~L. Feringa, ``Fine tuning of the
  rotary motion by structural modification in light-driven unidirectional
  molecular motors,'' {\em Journal of the American Chemical Society}, vol.~128,
  no.~15, pp.~5127--5135, 2006.
\newblock PMID: 16608348.

\bibitem{Cizek1966}
J.~Čížek, ``On the correlation problem in atomic and molecular systems.
  calculation of wavefunction components in ursell‐type expansion using
  quantum‐field theoretical methods,'' {\em The Journal of Chemical Physics},
  vol.~45, no.~11, pp.~4256--4266, 1966.

\bibitem{Bartlett2007}
R.~J. Bartlett and M.~Musia\l{}, ``Coupled-cluster theory in quantum
  chemistry,'' {\em Rev. Mod. Phys.}, vol.~79, pp.~291--352, Feb 2007.

\bibitem{Kohn1965}
W.~Kohn and L.~J. Sham, ``Self-consistent equations including exchange and
  correlation effects,'' {\em Phys. Rev.}, vol.~140, pp.~A1133--A1138, Nov
  1965.

\bibitem{Porezag1995}
D.~Porezag, T.~Frauenheim, T.~K\"ohler, G.~Seifert, and R.~Kaschner,
  ``Construction of tight-binding-like potentials on the basis of
  density-functional theory: Application to carbon,'' {\em Phys. Rev. B},
  vol.~51, pp.~12947--12957, May 1995.

\bibitem{Elstner1998}
M.~Elstner, D.~Porezag, G.~Jungnickel, J.~Elsner, M.~Haugk, T.~Frauenheim,
  S.~Suhai, and G.~Seifert, ``Self-consistent-charge density-functional
  tight-binding method for simulations of complex materials properties,'' {\em
  Phys. Rev. B}, vol.~58, pp.~7260--7268, Sep 1998.

\bibitem{Kulichenko2021}
M.~Kulichenko, J.~S. Smith, B.~Nebgen, Y.~W. Li, N.~Fedik, A.~I. Boldyrev,
  N.~Lubbers, K.~Barros, and S.~Tretiak, ``The rise of neural networks for
  materials and chemical dynamics,'' {\em The Journal of Physical Chemistry
  Letters}, vol.~12, no.~26, pp.~6227--6243, 2021.

\bibitem{Zubatiuk2021}
T.~Zubatiuk and O.~Isayev, ``Development of multimodal machine learning
  potentials: Toward a physics-aware artificial intelligence,'' {\em Accounts
  of Chemical Research}, vol.~54, no.~7, pp.~1575--1585, 2021.
\newblock PMID: 33715355.

\bibitem{Dral2020}
P.~O. Dral, ``Quantum chemistry in the age of machine learning,'' {\em The
  Journal of Physical Chemistry Letters}, vol.~11, no.~6, pp.~2336--2347, 2020.
\newblock PMID: 32125858.

\bibitem{Dellago1998}
C.~Dellago, P.~G. Bolhuis, and D.~Chandler, ``Efficient transition path
  sampling: Application to lennard-jones cluster rearrangements,'' {\em The
  Journal of Chemical Physics}, vol.~108, no.~22, pp.~9236--9245, 1998.

\bibitem{Bolhuis1998}
P.~G.~Bolhuis, C.~Dellago, and D.~Chandler, ``Sampling ensembles of
  deterministic transition pathways,'' {\em Faraday Discuss.}, vol.~110,
  pp.~421--436, 1998.

\bibitem{Metropolis1953}
N.~Metropolis, A.~W. Rosenbluth, M.~N. Rosenbluth, A.~H. Teller, and E.~Teller,
  ``Equation of state calculations by fast computing machines,'' {\em The
  Journal of Chemical Physics}, vol.~21, no.~6, pp.~1087--1092, 1953.

\bibitem{Crooks2001}
G.~E. Crooks and D.~Chandler, ``Efficient transition path sampling for
  nonequilibrium stochastic dynamics,'' {\em Phys. Rev. E}, vol.~64, p.~026109,
  Jul 2001.

\bibitem{Lubbers2018}
N.~Lubbers, J.~S. Smith, and K.~Barros, ``Hierarchical modeling of molecular
  energies using a deep neural network,'' {\em The Journal of Chemical
  Physics}, vol.~148, no.~24, p.~241715, 2018.

\bibitem{Stewart1989}
J.~J.~P. Stewart, ``Optimization of parameters for semiempirical methods i.
  method,'' {\em Journal of Computational Chemistry}, vol.~10, no.~2,
  pp.~209--220, 1989.

\bibitem{Zhou2022}
G.~Zhou, N.~Lubbers, K.~Barros, S.~Tretiak, and B.~Nebgen, ``Deep learning of
  dynamically responsive chemical hamiltonians with semiempirical quantum
  mechanics,'' {\em Proceedings of the National Academy of Sciences}, vol.~119,
  no.~27, p.~e2120333119, 2022.

\bibitem{Roke2018}
D.~Roke, S.~J. Wezenberg, and B.~L. Feringa, ``Molecular rotary motors:
  Unidirectional motion around double bonds,'' {\em Proceedings of the National
  Academy of Sciences}, vol.~115, no.~38, pp.~9423--9431, 2018.

\bibitem{Feringa2017}
B.~L. Feringa, ``The art of building small: From molecular switches to motors
  (nobel lecture),'' {\em Angewandte Chemie International Edition}, vol.~56,
  no.~37, pp.~11060--11078, 2017.

\bibitem{terWiel2005}
M.~K.~J. ter Wiel, R.~A. van Delden, A.~Meetsma, and B.~L. Feringa,
  ``Light-driven molecular motors: Stepwise thermal helix inversion during
  unidirectional rotation of sterically overcrowded biphenanthrylidenes,'' {\em
  Journal of the American Chemical Society}, vol.~127, no.~41,
  pp.~14208--14222, 2005.
\newblock PMID: 16218615.

\bibitem{Cnossen2014}
A.~Cnossen, J.~C.~M. Kistemaker, T.~Kojima, and B.~L. Feringa, ``Structural
  dynamics of overcrowded alkene-based molecular motors during thermal
  isomerization,'' {\em The Journal of Organic Chemistry}, vol.~79, no.~3,
  pp.~927--935, 2014.
\newblock PMID: 24410498.

\bibitem{Koumura2000}
N.~Koumura, E.~M. Geertsema, A.~Meetsma, and B.~L. Feringa, ``Light-driven
  molecular rotor: Unidirectional rotation controlled by a single stereogenic
  center,'' {\em Journal of the American Chemical Society}, vol.~122, no.~48,
  pp.~12005--12006, 2000.

\bibitem{vanDelden2003}
R.~A. van Delden, N.~Koumura, A.~Schoevaars, A.~Meetsma, and B.~L. Feringa, ``A
  donor–acceptor substituted molecular motor: unidirectional rotation driven
  by visible light,'' {\em Org. Biomol. Chem.}, vol.~1, pp.~33--35, 2003.

\bibitem{Pijper2007}
D.~Pijper and B.~Feringa, ``Molecular transmission: Controlling the twist sense
  of a helical polymer with a single light-driven molecular motor,'' {\em
  Angewandte Chemie International Edition}, vol.~46, no.~20, pp.~3693--3696,
  2007.

\bibitem{Behler2007}
J.~Behler and M.~Parrinello, ``Generalized neural-network representation of
  high-dimensional potential-energy surfaces,'' {\em Phys. Rev. Lett.},
  vol.~98, p.~146401, Apr 2007.

\bibitem{Zhou2020}
G.~Zhou, B.~Nebgen, N.~Lubbers, W.~Malone, A.~M.~N. Niklasson, and S.~Tretiak,
  ``Graphics processing unit-accelerated semiempirical born oppenheimer
  molecular dynamics using pytorch,'' {\em Journal of Chemical Theory and
  Computation}, vol.~16, no.~8, pp.~4951--4962, 2020.
\newblock PMID: 32609513.

\bibitem{Neese2020}
F.~Neese, F.~Wennmohs, U.~Becker, and C.~Riplinger, ``The orca quantum
  chemistry program package,'' {\em The Journal of Chemical Physics}, vol.~152,
  no.~22, p.~224108, 2020.

\bibitem{Becke1988}
A.~D. Becke, ``Density-functional exchange-energy approximation with correct
  asymptotic behavior,'' {\em Phys. Rev. A}, vol.~38, pp.~3098--3100, Sep 1988.

\bibitem{Perdew1986}
J.~P. Perdew, ``Density-functional approximation for the correlation energy of
  the inhomogeneous electron gas,'' {\em Phys. Rev. B}, vol.~33,
  pp.~8822--8824, Jun 1986.

\bibitem{Smith2018}
J.~S. Smith, B.~Nebgen, N.~Lubbers, O.~Isayev, and A.~E. Roitberg, ``Less is
  more: Sampling chemical space with active learning,'' {\em The Journal of
  Chemical Physics}, vol.~148, no.~24, p.~241733, 2018.

\bibitem{Conn2000}
A.~R. Conn, N.~I.~M. Gould, and P.~L. Toint, {\em Trust Region Methods}.
\newblock Society for Industrial and Applied Mathematics, 2000.

\end{thebibliography}
\bibliographystyle{ieeetr}
\end{document}